# Deriving AOC C-Models from D&V Languages for Single- or Multi-Threaded Execution Using C or C++

Tobias STRAUCH
R&D EDAptix
Munich, Germany
tobias@edaptix.com

**Abstract**

The C language is getting more and more popular as a design and verification language (DVL). SystemC, ParC [1] and Cx [2] are based on C. C-models of the design and verification environment can also be generated from new DVLs (e.g. Chisel [3]) or classical DVLs such as VHDL or Verilog. The execution of these models is usually license free and presumably faster than their alternative counterparts (simulators). This paper proposes activity-dependent, ordered, cycle-accurate (AOC) C-models to speed up simulation time. It compares the results with alternative concepts. The paper also examines the execution of the AOC C-model on a multithreaded processor environment.

**1. Introduction**

C based design and verification languages (DVL) have made an significant impact on the overall design process throughout the last decades. What has been dominated by classical languages like VHDL and Verilog (HDL) is now challenged by a fundamentally different approach. The C language is used to model the design and verification environment. For that the design or the testbench are either written in a syntax that is an extension to C, or the model is automatically translated to C from other languages like VHDL and Verilog.

System level design in C++ is proposed by Verkest et al. in [4]. A language that can be seen as an extension to C is for example SystemC [5]. The code can be directly compiled into an executable for simulation and it can be used for synthesis. Speeding up SystemC simulation is shown by Naguib et al. in [6]. A C-model is also used as an intermediate format in the design and verification flow. Design and testbenches written in languages like Cx [2], Chisel [3], VHDL or Verilog are translated into C, which can then be compiled with standard C compilers. An examples for tools converting Verilog to C is the verilator [7], for converting Verilog into an intermediate format iverilog [8], and for converting VHDL to machine code GHDL [9]. It is also proposed to co-simulate design elements in C and other languages. Bombana et al. demonstrate VHDL and C level cosimulation in [10] and Patel et al. evaluate on cosimulation of Bluespec and C based design elements in [11].

C-models can be cycle or timing accurate representations of the design and test behavior. This is true for most DVLs. In this paper it is assumed, that synthesis does not consider timing relevant aspects (like "delays" for instance) and that the design under test (DUT), which is used for synthesis, is modeled cycle accurately. A cycle (and not timing) accurate description of the DUT can be seen as good design practice, regardless which language is used. A classical example cycle accurate simulation is Hornet, a cycle level multicore simulator proposed by Ren et al. in [12]. Cycle based simulation using decision diagrams (DD) is discussed by Ubar et al. in [13] and based on reduced colored Petri net (RCPN) by Reshadi et al. in [14].

In this paper an activity-dependent, ordered and cycle-accurate (AOC) C-model of the DUT is proposed. Synthesis techniques are used to convert the RTL design into an elaborated representation. A clock tree analysis enables a cycle accurate simulation of the DUT. The proposed method allows an activity-dependent calculation of different design elements within individual clock domains. The model can also be executed on a multiprocessor system or on a multithreaded processor.

Section 2 describes the translation process of a DUT into a cycle-accurate C-model representation. In section 3 the algorithm is enhanced to support AOC C-models. How the model can be improved to support a multithreaded processor is shown in section 4. Section 5 describes how the AOC C-model can be combined with other verification relevant aspects. The proposed model is then compared to alternative concepts (section 6).

**2. C-Model Generation**

This section describes the C-model generation process. An algorithm is outlined in Figure 1, which supports the process of translating a design from any common language like Verilog or VHDL into a C-model.

1) Parsing source code
2) Hierarchy generation and parameter passing
3) Function and procedure enrollment
4) Variable unification and ordering
5) Signal and register detection
6) Clock tree detection and dependencies
7) Register and signal dependencies
8) Design graph optimizations
9) C code dumping

Figure 1: Algorithm for RTL to C-model conversion.

After parsing the source code, the design hierarchy is elaborated. During this step, parameter must be passed and generate statements must be considered. Step 3 covers the enrollment of functions, tasks and procedures. For both coding languages (VHDL and Verilog) a variable unification and ordering (step 4) within a single process must be done. After this initial phase, signals and registers need to be identified (step 5). The register detection leads to the step of clock line elaboration for each register. This information is then collected to group registers to individual clock domains and the dependencies of the clock domains itself (e.g. internal generated clocks, step 6). The aspect of using a sensitivity list becomes obsolete. Instead a register and signal ordering based on their dependencies takes place (step 7) and the resulting desing graph is further optimizes (step 8). Finally the design is dumped as C code (step 9).

The conversion algorithm (Figure 1) is common to most HDL-to-C translation tools. After parsing and elaborating the design, the database models the design in a design language independent format. In some alternative design flows, the design is already available in a C-model like fashion and the conversion and mapping steps are less complex. From step 6 onwards, the different language specific aspects of the source code become irrelevant. The mapping of each RTL statement for the Verilog and VHDL languages into C statements is listed in Table 1.

Table 1. VHDL/Verilog syntax mapping

| RTL | VHDL | Verilog | C |
|---|---|---|---|
| if | if the else | … ? … : … / if else | if () {} else {} |
| case | case (sel) when | case (sel) | if () {} else {} |
| math | a + b, -, *, … | a + b, -, *, … | +, -, *, ... |
| comb | not, and, or, … | ~, &, \|, … | !, &, \|, ... |
| unary |  | &a, \|a, ^a | !, &, \|, ... |
| mux | a(i) | a[i] | a[i] |
| demux | a(i) <= | a[i] <= | a[i] |
| shift | shl, shr | >>, << | >>, << |

It is important for the execution speed, how the design is represented when simulated. Therefore the steps 8 "Design graph optimization" and 9 "C code dumping" have a huge impact on the simulation performance of the C-model. The next section outlines various aspects of the design graph optimization and modeling aspects.

## 3. AOC C-Model Generation

The activity dependent, ordered, cycle accurate (AOC) C-model generation is discussed in this section. To a certain extend, almost all alternative models are AOC models. Some values (registers) are only calculated at a certain (clock) event (activity dependent), values must be calculated based on an ordered list (otherwise it will get very complicated if not impossible) and almost all models are cycle accurate. Nevertheless, different design representation aspects and different design graph optimization methods can lead to different execution speeds. Numbers will be shown in the result section. Figure 2 lists the various aspects which are discussed in this section.

1) N level signal modeling
2) Multidimensional types and type size
3) Direct computations vs. function calls
4) Design flattening and optimizations
5) Clock and output domain modeling
6) Register ordering
7) Wire ordering
8) Activity dependent signal ordering

Figure. 2: Design Graph Optimization Methods and Design Modeling Aspects

*3.1 Definitions*

Given is a set of inputs $I$, outputs $O$, sequential elements $R$ and a directed graph $G$ of combinatorial elements $C$ and wires $W$. The simplest form of a combinatorial element $c \in C$ is an assignment (buffer). An $c$ input ($ci$) can be an input $i \in I$, register $r \in R$ or wire $w \in W$. A $c$ output ($co$) can be an output $o \in O$, register $r \in R$ or wire $w \in W$. All $w$ have one driving combinatorial element $c$. All $c$ and $w$ build a directed graph without functional loops. A signal $s \in S$ can be an $i$, $o$, $r$, or $w$ ({$I, O, R, W$} $\in S$). A register $r$ can be an event or level sensitive sequential element.

A $cr \in CR$ is a clock root and $CR$ a list of all $cr$ of the design. A $cd \in CD$ is a set of registers with identical $cr$. $CD$ is a list of all $cd$ in the design. The register cone input list $rcil(r)$ is register specific and is a complete list of register and input signals which drive the directed tree *tr* of combinatorial elements (*c*) with (*r*) at its root. A primary input *pi* can be an input *i* or the output of a sequential element *r*. A primary output *po* can be a register input r or an output *o*. All *po* with the same clock root *cr* are grouped to a primary output domain *pod(cr)*. The *POD* lists all *pod(cr)* of the design.

*3.2 N-level Signal Modeling*

For the optimization techniques discussed now, it is assumed that C variables of the standard type "unsigned" generate faster execution models than their comparable representation as a specific class. In the proposed AOC model generation, a 2-value representation of *s* is therefore default, unless specified otherwise. Assuming the signal s0 is an 8 bit wide bus and should only simulate {0, 1}, then s0 can be of type "unsigned". If s0 should simulate more than 2 values {0, 1, X, Z, ...}, then s0 must be represented by a specific signal-class.

*3.3 Multidimensional Types and Type Size*

The different signal types are elaborated and serialized. Let cw be the bit width of the type "unsigned" of the target architecture for the model execution. A classical cw of a processor architectures is 32 or 64. If (serialized) types have more than cw bits, then the C representation is a two dimensional array of the serialized type. An exception to this rule is a 2-dimensional array type with less than cw bits per dimension. In this case the type is also modeled as a "2-dimensional unsigned array" but not serialized. If the signal must be modeled as a signal-class, then the class can use a serialized or a dynamic representation.

*3.4 Direct Computations vs. Function Calls*

Each combinatorial element *c* should be modeled as a direct computation "a = b & c;" and not as a function call "a = AND(b, c);". If at lease one signal is a signal-class, then a function call "AND(a, b, c);" is required. Functions must be provided to convert signals of type unsigned to or from a signal-class.

*3.5 Design Flattening and Optimizations*

The design hierarchy is removed by flattening the design. Signal names are modified to guarantee the uniqueness of the signal. For a better readability (debugging), the hierarchical instantiation names are typically added as a prefix to the signal name (e.g. topi_subsystem1i_cpui_executei_pc) but any other method/prefix to uniquify the signals is applicable.

The design builds a directed graph *G* of combinatorial elements *C* and signals *S*. Constant values (e.g. a signal is driven by a constant value "s <= 1'b0;") are propagated through G and all $c \epsilon C$ and $s \epsilon S$ that become irrelevant are removed. Also direct assignment pairs "s2 <= s1; s1 <= s0;" are simplified "s2 <= s0;" and the irrelevant entries in the design database (*c*, *s*) are removed. Most of these direct assignment pairs result from design flattening.

*3.6 Clock and Output Domain Modeling*

After the register identification step, the clock (or enable) input of each register (*r*) is traced back to its clock root (*cr*). Clock roots (*cr*) can be inputs (*i*) to the design, outputs of combinatorial logic (*w*) or registers (*r*). The clock roots which are design inputs become independent driver of their individual clock domain (*cd*). Clock roots which are latch or register outputs or outputs of logic cones are drivers of clock domains, which dependent on other clock domains (*cd*).

All outputs $o \in O$ are automatically grouped to the output domain *OD* (= *O*) $\in$ *POD*. They are independent of any clock (or enable) event and their value must be calculated whenever one of the *pi* of their directed tree *tr(o)* has changed its value.

*3.7 Register Ordering*

The register list of each clock domain (*cd*) must be ordered, based on their interdependency. For that the register cone input list (*rcil(r)*) is generated for each register. Figure 3 shows an example.

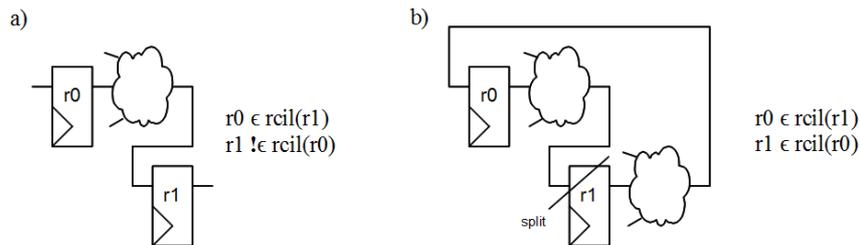

Figure 3: Simple Ordering and Ordering Using Split

The ordering of all *r* of one *cd* is trivial when their *rcil(r)* is used. The logic cones of Figure 3a can easily be ordered and calculated. The registers of Figure 3b depend on each other and no clear order can be found. At least one register must therefore be splitted into its output value and a pre-register value that is calculated first. All relevant *r* then take over their pre-register value as an output value at the end of the modeling task of a clock domain.

*3.8 Wire Ordering*

The directed tree *tr(po)* for each primary output *po* of a primary output domain *pod(cr)* must be modeled. A *po* can either be a *r* or an *o*. Therefore the modeling of all wires $w \in tr(po)$ must be ordered based on their interdependency. A *w* gets the attribute $w^{cal}$ once it has been added to the list of calculated *w WL*. Then it can be defined, that the value of a *w* can be calculated when the inputs of the associated combinatorial logic element *c* are *r*, *pi*, or $w^{cal}$. This constraint allows an ordering of all $w \in tr(po)$. The *po* value can be calculated when all $w \in tr(po)$ are $w^{cal}$. Each $w^{cal}$ holds its attribute until all *po* of a *POD* are calculated. Therefore each relevant *w* of a *pod(cr)* is only calculated once.

*3.9 Activity Dependent Signal Ordering*

An special modeling technique is the activity dependent signal ordering (ADSO). A combinatorial element *c* changes its output value *co* only when at least one of their inputs *ci* has changed. Classical HDLs like VHDL and Verilog support this fact by using a sensitivity list. Figure 4 shows how ADSO is implemented in an AOC model.

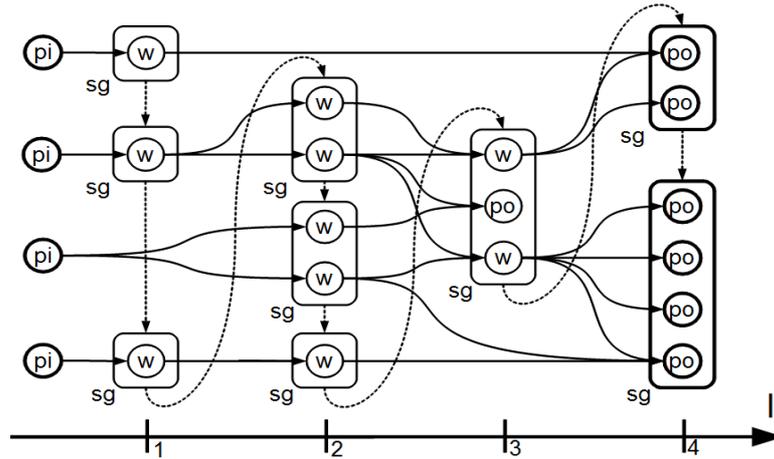

Figure 4: Implementing Activity Dependent Signal Ordering

The ADSO evaluation algorithm is based on two steps. In the first step, all outputs of combinatorial elements *co* (= {*w, r, po*}) are placed on different levels *l* based on the following rules. **Rule 1** says, that each *co* must be placed on the lowest possible level. **Rule 2** says, that a *co* which depends on a list of *ci* (= {*w, pi*}) *lci* must be placed on a higher level than all *w* of *lci*. In the second step, all *co* on a level *l* are grouped to signal groups *sg(l)* based on the following rule. **Rule 3** says, that all *co* on one *l* which depend on at least one identical wire *w* are grouped to a *sg(l)*. The ordered signal list *OSL* (dotted line in Figure 4) holds all *sg(l)* ordered by their individual level *l* assignment.

The ADSO execution algorithm adds the active attribute $sg^{act}$ to each *sg*. If a *ci* changes its value during execution, then the following is applied. **Rule 4** says, that the $sg^{act}$ is set for the *sg* for which one of the *co* has the *ci*. All *sg* are evaluated based on the *OSL*. All *co* in an *sg* are only calculated, if the $sg^{act}$ attribute is set. After the execution of an *sg*, the $sg^{act}$ attribute is cleared again.

## 4. Multithreaded Execution of AOC Models

When an AOC model should be executed on a multithreaded processor (or multiprocessor) environment then the AODS algorithm must be enhanced by the following steps. In step 1, the combinatorial elements *c*, wires *w* and output *o* of the output domain *OD* are added to each individual clock domain *cd*, which are connected to the *po* of this *cd*. The resulting clock domain is then called *cdo*. In step 2, the elements of a *cdo* are partitioned to be executed on individual threads *td*. The number of maximal treads *tdmax* must be defined. Each individual *po* of the *cdo* is uniquified and an individual ordered signal list *OSL* is generated. Therefore, *cdo* elements are duplicated if they are elements of individual *OSL*. The *OSL* must be then merged as long as the number of *OSL* is greater than *tdmax* based on the following rule. **Rule 5** says, that these two *OSL* out of all *OSL* are merged, which share the highest number of *cdo* elements. Figure 5 shows an example of 3 *cdo* and 2 *td*/*OSL*.

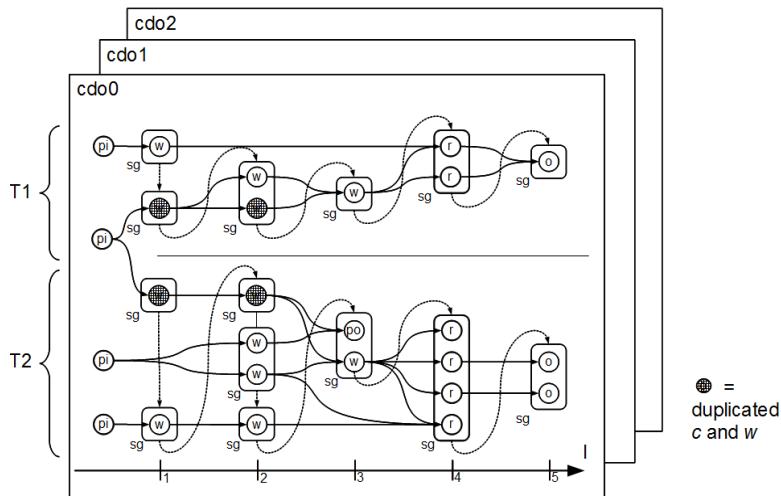

Figure 5: Enhanced Model for Multithreaded Execution

The memory model is critical for the program performance, especially when executed on a multithreaded or multicore system. The "OpenCL Memory Model" [15] is shown in Figure 6. It is used to demonstrate the memory usage of AOC models. Wires $w$ do have a very short lifetime and do not need to be shared among multiple threads. They are therefore stored in the "Private Memory" (Figure 6) of a work-item (= thread). Registers $r$ are stored throughout the program runtime and most of them must be shared among multiple work-items. They are located in the "Local Memory" (Figure 6) of a workgroup and become primary inputs $pi$ in the next cycle. Outputs $o$ are calculated for each cycle and stored in the "Constant Memory" of the "Compute Device" (Figure 6). Inputs $i$ are also stored in the "Constant Memory" to be used by the work-items as $pi$.

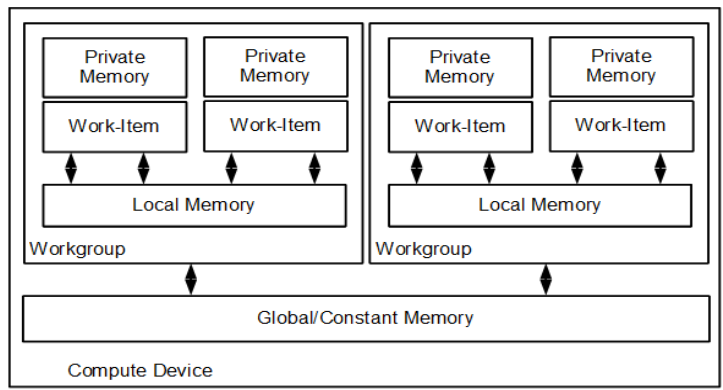

Figure 6: OpenCL Memory Model of the Compute Device

In the proposed *OSL* based modeling technique, wires $w$ only have a very short lifetime. Two steps can be made to improve the execution speed. In step 1, the following observations and definitions are made. The order of $w$ calculation within one level can be freely selected. It is defined that a wire pair *wp* has a first wire *fw* and a second wire *sw*, whereas the *sw* depends on the *fw*. The *wp* with the *fw* and the *sw* on consecutive levels are added to a level specific wire pair list *WPL(l)*. The *wp* of which the *fw* is only used by the *sw* are listed last in the *WPL(l)*. The following rule is defined to keep the values of $w$ in the register file of a processor during execution. **Rule 6** says, that for each *wp* of the ordered *WPL(l)* the *fw* is added at the end and the *sw* is added at the beginning of the $w$ calculation of each level. This increases the chances that the compiler avoids memory accesses and keeps the temporary values of $w$ in the register file.

In step 2, it is tried to avoid further time consuming memory access by reusing cache entries for multiple wire calculations. To achieve this, a list of placeholders *PHL* and its maximal size (cache size) *phmax* is defined. **Rule 7** says, that a *w* is assigned to a *PHL* entry, if its value is used in the remaining execution of the SGL. If the *w* is not needed anymore, than its *PHL* entry can be used by another *w*. This rule and *phmax* must already be considered when two *OSL* are merged to a single one (rule 5). Rule 7 can help the compiler to store values in a local cache. The parameters *tdmax* and *phmax* are system specific.

Seven rules have been defined throughout the last two sections. They define the transformation process of the device under test (DUT) - which can be seen as a directed graph G defined in section 3.1 - into an activity-dependent, ordered and cycle-accurate (AOC) C-model.

**5. Testbenches**

This paper discussed so far how synthesizeable HDL code can be transformed into AOC C-models. An extended flow can be used to transfer testbenches (non-synthesize-able code) into timing accurate representations in C++ format. Both models can then be linked to execute the same simulation process as known from HDL simulators.

The additional aspects of this flow are the DUT definition and the sequential process identification. Sequential processes are HDL statements, which are time consuming due to timing related statements (example: "wait for 10 ns;") or conditional statements (example: "wait on <signal>;"). These processes cannot be converted into cycle accurate models, they need to be modeled timing accurate. The resulting model has an condition checker and an event scheduler. Both reflect the entries of the sequential processes. The flow for this methodology is outlined in Figure 7.

1) parsing source code
2) parameter passing and hierarchy
3) function and procedure enrollment
4) variable uniquification
5) DUT definition
6) sequential process identification
7) signal and register detection
8) clock tree detection and dependencies
9) register and signal logic cone conversion
10) register and signal dependencies
11) C++ code optimization and dumping
12) condition bag dumping
13) event scheduler dumping

Figure 7: Algorithm for HDL Testbench to Timing Accurate Modeling

A process in VHDL or Verilog is defined as a time-consuming process (TCP) when it is not within the DUT hierarchy and when the keyword "wait" is used within this process. A TCP is partitioned into individual list of assignments, based on the different wait statement. A list of assignments is continuously executed until a "wait" statement is reached. This event is added to the event list in case of a "wait for" statement, and to a the condition list in case of a "wait on" statement. The execution of the assignment list of the TCP is continued, once the simulation time reaches the event in the event list or when the condition in the condition list is true. The conditions in the condition list are checked very simulation step.

## 6. PSL, Waveform and MatLab

*6.1 PSL*

The property specification language (PSL) can be used for checking design behavior during verification. The language supports different flavors (VHDL, Verilog) and can be part of the design RTL source code itself. The language is more and more used already on C/SystemC level or higher level of abstraction as proposed by Habibi et al. in [16]. It is therefore important to incorporate the structure into AOC C-models. Obereder et al. describe in [17], how PSL can be converted into synthesizable HDL code. This approach can be used and the resulting synthesizable HDL code can be converted into AOC C-models.

*6.2 Waveform*

Just like common HDL simulators, AOC C-models can dump simulation waveforms as well. Signals can be defined manually, by script or by HDL syntax (example: Verilog). The AOC C-model then dumps a cycle accurate VCD file during execution, which can then be viewed by a standard waveform viewer.

*6.3 Running AOC C-Models in Matlab*

AOC C-models can also be executed in a Matlab [18] based environment. For that the C code must be compiled into an S-function. It can then be co-simulated together with other Matlab based simulation components. This is very useful for accelerators or custom DSPs designed in HDL. They can then be simulated and verified in a much more flexible simulation environment than classical HDLs can offer.

## 7. Results

This section compares the execution speed of AOC C-models to alternative concepts. Different testcases are used to measure the individual runtimes. The Verilog version of the OpenRISC SoC was taken from [19] and a testcase was added. The verilator [7] and iverilog [8] runtime was then compared to the runtime of the AOC C-model, which was automatically generated from the Verilog source code. A single stage RISCV32IM processor [20] with a lengthy testcase was developed in VHDL and SystemC. A Chisel version was taken from [21]. Their runtime was then compared to the runtime of the AOC C-model, which was automatically generated from the VHDL source code. A license for a standard simulator was not available. This is why a comparison number of verilator vs. VCS was taken from [7] and added to the list as a VCS vs. AOC C-model relative runtime entry, considering the fact, that the AOC C-mode is about 5.09 times faster than the verilator execution. The numbers are based on tests executed on a single processor system. Table 2 and Figure 8 show the results. The AOC C-models are always the 100% reference runtime.

**Table 2.** Relative runtime compared to AOC C-models.

|  | verilator | iverilog | GHDL | SystemC | Chisel | VCS |
|---|---|---|---|---|---|---|
| Relative Runtime | 5.09 | 18.39 | 11.44 | 7.00 | 3.91 | 20.6 |
| AOC C-model | 1 | 1 | 1 | 1 | 1 | 1 |

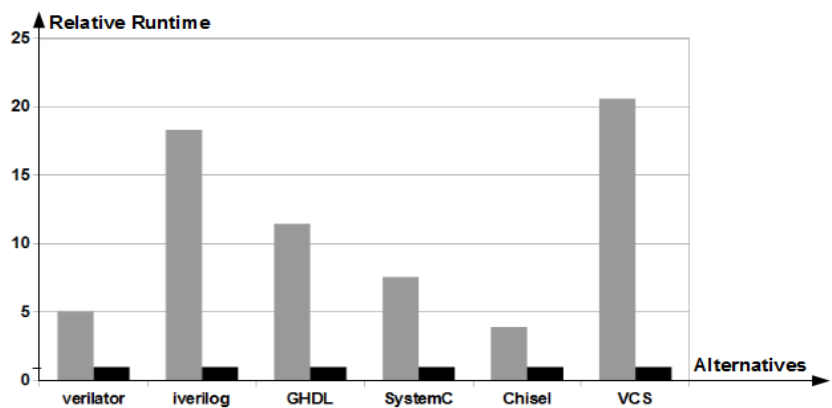

Figure 8: Relative runtime comparison of the AOC C-model and alternatives.

The numbers show that the AOC C-models are always faster than any other known C-model (iverilog, ….) based alternative. The compile time of the individual C-models are almost the same and their comparison can be neglected.

## 8. Conclusion

This paper introduced activity dependent cycle accurate (AOC) C-models. Multiple improvement steps can be applied to successively decrease the execution time. The main improvements are the activity dependent solving of combinatorial logic equations and their execution based on an extracted signal order. Tests show that they have a faster execution speed than pure cycle accurate C-models or HDL simulators.

This paper also shows that the proposed AOC C-model fits nicely into a multiprocessor system .The current research concentrates on generating AOC OpenCL-models for multiprocessor or multithreading processor systems.